\documentclass [aps,prb,10pt,twocolumn,showpacs,superscriptaddress]{revtex4-1}
\usepackage[english]{babel}
\usepackage{graphicx}
\usepackage{amsmath,amssymb}
\usepackage[utf8]{inputenc}
\usepackage{natbib}
\usepackage{units}
\usepackage{hyperref}
\usepackage{color,soul}
\usepackage{url}
\hypersetup{colorlinks=true,citecolor={blue},linkcolor={blue},urlcolor={blue}}
\usepackage{balance}

\begin{document}
\title{Half-Skyrmion Spin Textures In Polariton Microcavities}

\date{\today}

\author{P. Cilibrizzi}
\email[correspondence address:~]{pasquale.cilibrizzi@soton.ac.uk}
\affiliation{School of Physics and Astronomy, University of Southampton, Southampton, SO17 1BJ, United Kingdom}
\author{H. Sigurdsson}
\affiliation{Division of Physics and Applied Physics, School of Physical and Mathematical Sciences, Nanyang Technological University 637371, Singapore}
\affiliation{Science Institute, University of Iceland, Dunhagi-3, IS-107 Reykjavik, Iceland}
\author{T.C.H. Liew}
\affiliation{Division of Physics and Applied Physics, School of Physical and Mathematical Sciences, Nanyang Technological University 637371, Singapore}
\author{H. Ohadi}
\affiliation{School of Physics and Astronomy, University of Southampton, Southampton, SO17
1BJ, United Kingdom}
\author{A. Askitopoulos}
\affiliation{School of Physics and Astronomy, University of Southampton, Southampton, SO17
1BJ, United Kingdom}
\author{S. Brodbeck}
\affiliation{Technische Physik, Wilhelm-Conrad-R\"{o}ntgen-Research Center for Complex Material Systems,
Universit\"{a}t W\"{u}rzburg, Am Hubland, D-97074 W\"{u}rzburg, Germany}
\author{C. Schneider}
\affiliation{Technische Physik, Wilhelm-Conrad-R\"{o}ntgen-Research Center for Complex Material Systems,
Universit\"{a}t W\"{u}rzburg, Am Hubland, D-97074 W\"{u}rzburg, Germany}
\author{I. A. Shelykh}
\affiliation{Division of Physics and Applied Physics, School of Physical and Mathematical Sciences, Nanyang Technological University 637371, Singapore}
\affiliation{Science Institute, University of Iceland, Dunhagi-3, IS-107 Reykjavik, Iceland}
\author{S. H\"{o}fling}
\affiliation{Technische Physik, Wilhelm-Conrad-R\"{o}ntgen-Research Center for Complex Material Systems,
Universit\"{a}t W\"{u}rzburg, Am Hubland, D-97074 W\"{u}rzburg, Germany}
\affiliation{SUPA, School of Physics and Astronomy, University of St Andrews, St Andrews, KY16 9SS, United Kingdom}
\author{J. Ruostekoski}
\affiliation{Mathematical Sciences, University of Southampton, Southampton SO17 1BJ, United Kingdom}
\author{P. Lagoudakis}
\affiliation{School of Physics and Astronomy, University of Southampton, Southampton, SO17
1BJ, United Kingdom}

\begin{abstract}
We study the polarization dynamics of a spatially expanding polariton condensate under nonresonant linearly polarized optical excitation. The spatially and temporally resolved polariton emission reveals the formation of non-trivial spin textures in the form of a quadruplet polarization pattern both in the linear and circular Stokes parameters, and an octuplet in the diagonal Stokes parameter. The continuous rotation of the polariton pseudospin vector through the condensate due to TE-TM splitting exhibits an ordered pattern of half-skyrmions associated with a half-integer topological number. A theoretical model based on a driven-dissipative Gross-Pitaevskii equation coupled with an exciton reservoir describes the dynamics of the nontrivial spin textures through the optical spin-Hall effect.
\end{abstract}

\maketitle

\section{Introduction}
\label{sec:1}

Skyrmions are \textit{non-singular} but topologically \textit{non-trivial} spin textures \cite{lewenstein_ultracold_book_2012}, identified by a winding number, known as the \textit{skyrmion number}, which corresponds to the number of times the spin vector continuously rotates across a finite region of space \cite{nagaosa_topological_skyrmion_2013}. In particular, they are \textit{non-singular} because their spin is always defined in each point of space (i.e., there are no singularities) and \textit{non-trivial} because they cannot be continuously transformed in a topologically trivial state (such as a ferromagnetic one, with all spins aligned in the same direction) and are hence relatively stable against perturbations \cite{ritz_spintronics:_2015}. This property makes skyrmions particularly attractive in the development of novel spintronics devices \cite{ritz_spintronics:_2015}. Although they were originally proposed by Skyrme in the field of nuclear physics \cite{skyrme_non-linear_1961}, skyrmions have recently received special attention in solid state systems, such as semiconductor quantum wells \cite{bernevig_quantum_2006} and ultrathin magnetic films \cite{yu_real-space_magnet_skyrmions_2010,Romming_Size_Shape_magnetic_skyrmions_2015}, due to their potential in future applications, such as low-power ultradense magnetic memories and logic devices \cite{fert_skyrmions_2013, Romming_Size_Shape_magnetic_skyrmions_2015}.
On a more fundamental level, three-dimensional (3D) skyrmions represent topological particle-like solitons in field theory, high-energy physics~\cite{manton-sutcliffe},
and in atomic superfluids~\cite{ruostekoski_prl_2001,al_khawaja_skyrmions_2001,battye_prl_2002,savage_prl_2003}. Moreover, two-dimensional (2D) skyrmions play a key role in the rotational properties of superfluid liquid $^3$He \cite{anderson_prl_1977,Mermin_vortices_1976}  and in atomic spinor Bose-Einstein condensates~\cite{mizushima_prl_2002,Leanhardt_coreless_vortex_2003,Leslie_2009,choi_prl_2012,Lovegrove_energetic_Stability_skyrmions_2014}, where they represent the vectorial counterpart of the quantized vortices of scalar superfluids and are usually referred to as coreless vortices, due to the absence of a vortex line singularity.
Recently, skyrmion spin textures were theoretically predicted also in indirect excitons \cite{Skyrmion_Indirect_exciton_2013} and exciton-polariton condensates \cite{flayac_transmutation_2013}. 
\\
\indent In this article, we report the formation and time evolution of 2D half-skyrmion spin textures in planar semiconductor microcavities, which are suitable systems for studying the fundamental properties of dissipative bosonic systems, such as exciton-polariton condensates \cite{Carusotto_Rev_Mod_Physics_2013}. Exciton-polaritons, or hereafter polaritons, are composite bosonic quasiparticles formed by the strong coupling between heavy hole excitons, confined in quantum wells, and the photonic mode of a planar semiconductor microcavity \cite{kavokin_microcavities_2007}. By increasing the polariton population above a threshold density, polaritons can macroscopically occupy the ground state of the dispersion and form a non-equilibrium BEC \cite{byrnes_exciton-polariton_2014}, characterized by an inversion-less amplification of the polariton emission \cite{Imamoglu_polariton_laser_1996, Deng_polariton_lasing_2003} and macroscopic coherence over hundreds of microns \cite{Nelsen_macroscopic_coherence_2013}. Moreover, being bosons, polaritons possess an integer spin with two possible projections of the angular momentum $(S_z=\pm1)$ on the structural growth axis $(z)$ of the microcavity, which correspond to the right and left circular polarization of the emitted photons. Superpositions of the $S_z=\pm1$ states give rise to the linear or elliptical polarization states of polaritons.
An important effect, which capitalizes on the spin of polaritons, is the optical spin Hall effect (OSHE). After its first theoretical prediction \cite{kavokin_optical_2005}, the OSHE has been observed in both strongly coupled \cite{leyder_observation_2007} and weakly coupled \cite{maragkou_optical_2011} microcavities. The effect is a consequence of the energy splitting between transverse-electric (TE) and transverse-magnetic (TM) polarized modes \cite{panzarini_exciton_1999}, which occurs naturally in microcavities and represents an effective spin-orbit coupling. The initial demonstration of the optical spin Hall effect relied on resonant Rayleigh scattering \cite{langbein_polarization_2007}, while spin currents were similarly generated using tightly focused laser spots in both resonant \cite{leyder_observation_2007, maragkou_optical_2011} and non-resonant configurations \cite{kammann_nonlinear_2012, Carlos_1Dtexture_2015}. The TE-TM splitting also leads to the generation of vortices via spin-to-orbital angular momentum conversion \cite{liew_excitation_2007,manni_spin--orbital_2011}, which was recently shown to be enhanced in tunable open microcavity structures \cite{Spin_texture_open_cavity_2015}.
\\
\indent In polariton microcavities, skyrmions were theoretically predicted under resonant excitation \cite{flayac_transmutation_2013}. Differently from Ref.~[\onlinecite{flayac_transmutation_2013}], we use a nonresonant excitation scheme to ensure that the original coherence of the laser is lost in the relaxation process \cite{byrnes_exciton-polariton_2014}. We create a polariton condensate with a pseudospin orientation defined by the polarization of the excitation beam \cite{Hamid_2012,Li_G_inchoerent_polarization_pump_2015}. As the condensate expands, polaritons propagate over macroscopic distances, whilst their pseudospin collectively precesses. The three pseudospin orientations correspond to the three Stokes parameters (Eq.~\ref{eq:Stokes_eq}) measured from analyzing the polarization of the emission. We record the formation of intricate spin textures in the spatial expansion of the polariton condensate. A quadruplet pattern is observed in the linear and circular Stokes parameters, while an octuplet is observed in the diagonal Stokes parameter. The formation of the observed spin textures is described in the framework of the driven-dissipative Gross-Pitaevskii equation through the optical spin Hall effect~\cite{kavokin_optical_2005}. The half-skyrmion spin textures are unequivocally identified by their topological charge, i.e., the skyrmion number, calculated here to the best of our knowledge for the first time in polariton condensates.
\\
The article is organized as follows. In Sec.~\ref{sec:2} we describe the experimental setup and the sample. In Sec.~\ref{sec:3} the theoretical model is presented. In Sec.~\ref{sec:4} we report the main experimental results (\ref{subsec:a}), discuss the skyrmion number (\ref{subsec:b}), and describe the physical mechanism behind the formation of the spin textures (\ref{subsec:c}). Conclusions and perspectives are reported in Sec.~\ref{sec:5}.
\begin{figure*}[!hbtp]
  \centering
		\includegraphics[scale=0.51]{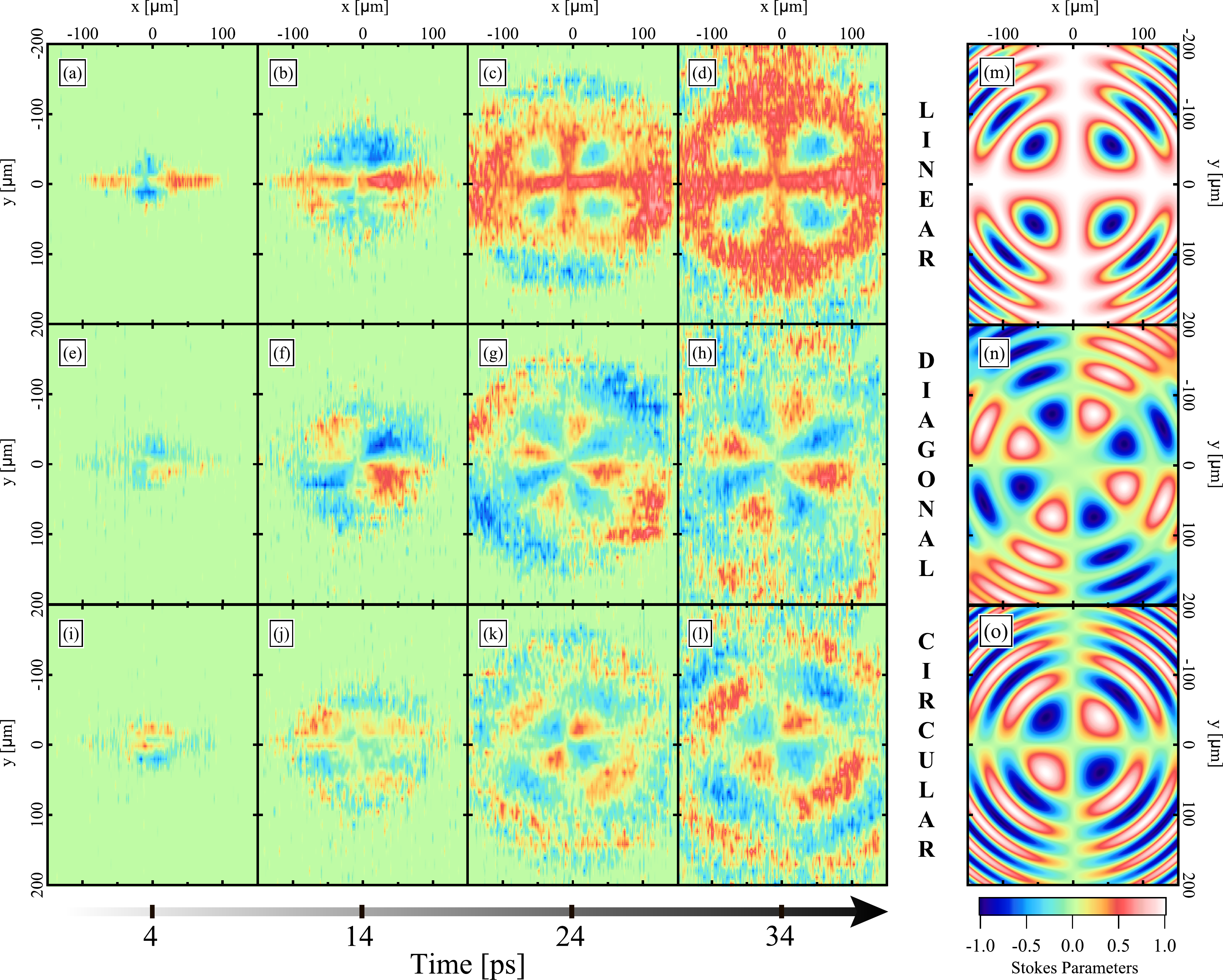}
	\caption{(Color online) Experimental real space linear (a-d), diagonal (e-h) and circular (i-l) Stokes parameters showing the formation dynamics of the polariton spin textures after pulsed optical excitation at $\unit[1.687]{eV}$. The excitation beam is horizontally polarized.  The zero time is defined at the PL onset as indicated in the spatial integrated intensity profiles in Figs.\ref{fig:3}(a-c). Theoretical real space linear (m), diagonal (n) and circular (o) Stokes parameters 34 ps after excitation with $P_+/P_- = 1$ and $P_+, \ P_- > 0$. The color scale is the same both for the experimental and simulated results.}
\label{fig:1}
\end{figure*}
%
%
%
%
\section{Sample and experiment}
\label{sec:2}

In this study we use a $\lambda/2$ AlGaAs/AlAs microcavity sample \cite{sampleA_2014} composed of 23 (27) pairs of AlGaAs/AlAs layers, forming the top (bottom) Distributed Bragg Reflectors (DBRs) and 4 triplets of $\unit[7]{nm}$ thick GaAs quantum wells (QWs) placed at the antinodes of the cavity electric field. The measured quality factor exceeds 9000 corresponding to a cavity photon lifetime of $\sim\unit[3.8]{ps}$. Strong coupling is obtained with a Rabi splitting energy of $\unit[14.5]{meV}$. The microcavity wedge allows one to choose the detuning between the exciton and the cavity mode. All the data presented here are recorded at a negative detuning of $\unit[-4.4]{meV}$. The sample is held in a cold-finger cryostat at a temperature of $T \approx 6$ K.
\\
To conduct the experiments we use the experimental setup schematically shown in Fig.\ref{fig:S1} of the Supplementary Information (SI). We use a mode-locked Ti-Sapphire pulsed laser to excite the sample at $\unit[1.687]{eV}$, corresponding to the first reflectivity minimum above the stopband of the DBR. The pulse width of the laser is $\sim\unit[180]{fs}$, at a repetition rate of $\unit[80]{MHz}$. The excitation beam is horizontally polarized and focused to a $\sim\unit[2]{~\mu m}$ at FWHM spot diameter using a 0.4 numerical aperture (NA) microscope objective. The average fluence of the excitation beam is kept at $\sim\unit[600]{\mu J/cm^2}$ throughout these measurements. 
Photoluminescence (PL) is collected in reflection geometry through the excitation microscope objective, analyzed by a polarimeter composed of a $\lambda/2$ or $\lambda/4$ plate and a linear polarizer and projected on the entrance slit of a streak camera with 2 ps temporal resolution.
\\
The spin of polaritons can be described in terms of the pseudospin ($\mathbf{S}$) formalism, in which the polarization of the light emitted from the cavity is characterized by the linear ($S_x$), diagonal ($S_y$) and circular ($S_z$) Stokes parameters, which corresponds to the following degree of polarizations:
\begin{equation}
S_{x}=\frac{I_{H}-I_{V}}{I_{H}+I_{V}},~ S_{y}=\frac{I_{D}-I_{A}}{I_{D}+I_{A}},~ S_{z}=\frac{I_{\sigma_+}-I_{\sigma_-}}{I_{\sigma_+}+I_{\sigma_-}},
\label{eq:Stokes_eq}
\end{equation}
where $I_{H,D,\sigma_+}$ and $I_{V,A,\sigma_-}$ are the measured intensities for the linear (Horizontal, Vertical, Diagonal, Anti-diagonal) and circular ($\sigma_+ , \sigma_-$) components. Thus, by measuring the polarization of the emitted light we record the polariton pseudospin state.
\\
The spatial polarization dynamics of the polariton expansion was time-resolved  using a tomography scanning technique. In this technique, the polarization analyzed PL intensity $I(x,y,t)$ is projected at the entrance slit of the streak camera. By using a motorized mirror, we scan the vertical direction, $y$, of the PL image and acquire $I(x,t)$ at different values of $y$. In this way, a 2D real space image $I(x,y,t)$ can be reconstructed as a function of time.
%
%
\section{Theoretical model}
\label{sec:3}

To model the spin dynamics of the polaritons BEC, we use a driven-dissipative Gross–Pitaevskii equation (2), describing the polariton field ($\Psi_{\pm}$), which is then coupled to an excitonic rate equation (3) describing a hot exciton reservoir ($\mathcal{N}_{\pm}$) generated by the nonresonant pump \cite{Wouters_NonResonant_GPE_2008}:
\begin{align} \notag
i \hbar \frac{d \Psi_{\pm}}{d t} &= \Big[ \hat{E}  - \frac{i \hbar }{2 \tau_p} + \alpha |\Psi_{\pm}|^2 + G  P_\pm (\mathbf{r},t) \\
&  + \Big( g_R + \frac{i \hbar r_c}{2} \Big) \mathcal{N}_{\pm} \Big] \Psi_{\pm} + \hat{H}_{\text{LT}} \Psi_{\mp},
\end{align}
\begin{equation}
\frac{d \mathcal{N}_{\pm}}{d t} = - \left( \frac{1}{\tau_x} + r_c |\Psi_{\pm}|^2 \right) \mathcal{N}_{\pm} +  P_\pm (\mathbf{r},t).
\end{equation}
Here the indices represent the spin up/down ($\pm$) basis. The coupled equations take into account a condensation rate $(r_c)$, corresponding to the rate at which excitons condense into polaritons and the energy blueshift of the polariton condensate due to interactions with excitons (with interaction strength $g_R$). The condensed polariton field obeys approximately a parabolic dispersion $\hat{E} = -\hbar^2 \nabla^2/2m^*$, where $m^*$ is the effective polariton mass. The polariton and exciton lifetimes are written $\tau_p$ and $\tau_x$ respectively. Same-spin polariton interaction strength is characterized by the parameter $\alpha$. We neglect interactions between polaritons with opposite spins, which are typically small in magnitude \cite{Vladimirova_2010} at energies far from the biexciton resonance \cite{Takemura_2014}. The exciton reservoir is driven by a Gaussian pump, $P_\pm(\mathbf{r},t)$, as described in section II. For example, a horizontally polarized pump would correspond to $\left\{P_+ = P_- \ : \ P_+, \ P_- \in \mathbb{R}^+\right\}$. The interaction constant $G$ represents an additional pump-induced shift which takes into account other excitonic contribution to the blueshift \cite{Wouters_NonResonant_GPE_2008}. 
\begin{figure*}[!hbtp]
  \centering
		\includegraphics[scale=0.14]{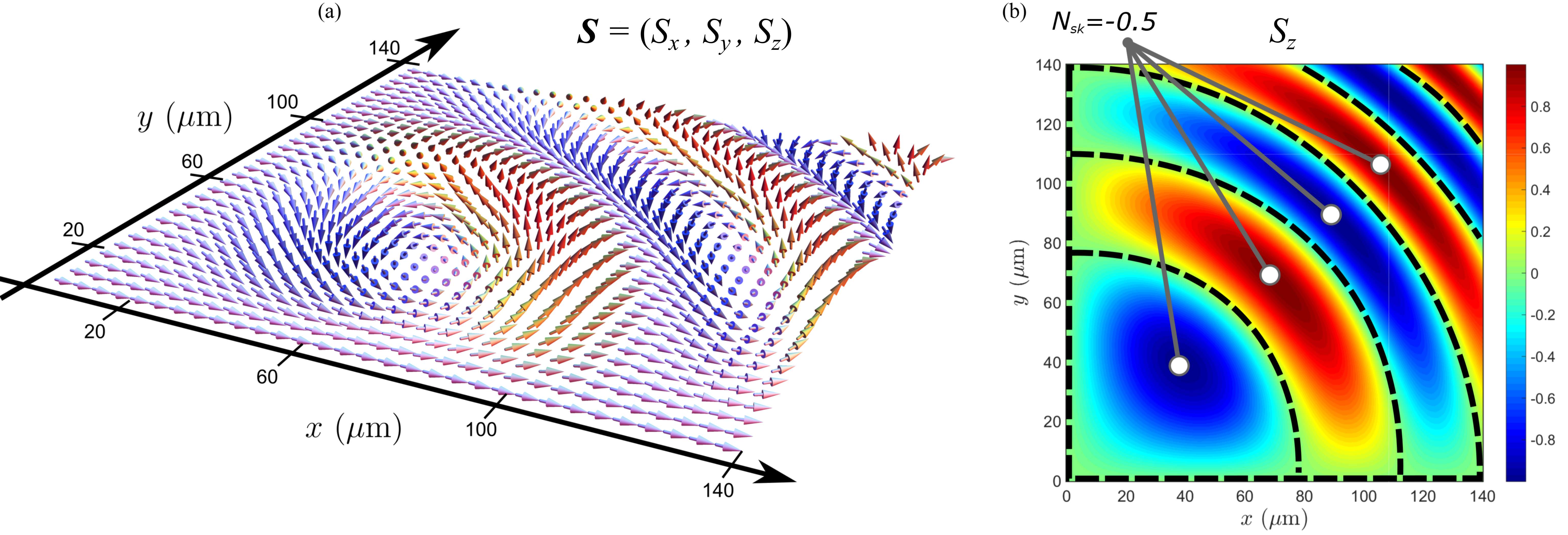}
	\caption{(Color online) (a) The vector field of the 2D polariton half-skyrmions, showing the rotation of the total pseudospin vector $\textbf{S}=(S_x,S_y,S_z)$ in the first quadrant of the microcavity $x$-$y$ plane (see Eq.1 of the SI). The different colors of the vectors refer to the different polarization domains. In particular, red and blue refer to the $\pm1$ circular polarizations of $S_z$ respectively (with opposite orientations along the $z$-axis), while pink in (a) and green in (b) to the linear and diagonal polarization components $S_x$, $S_y$ (i.e., the pseudospin lying in the $x$-$y$ plane). (b) Circular Stokes components $S_z$ showing the domains (circumscribed by the black dotted lines) where the skyrmion number $N_{sk}=-0.5$ has been calculated using Eq.~\ref{eq:Sk_number}.}
\label{fig:1b}
\end{figure*}
$\hat{H}_{\text{LT}}$ is the TE-TM splitting which mixes the spins of the polaritons:
\begin{equation}
\hat{H}_{\text{LT}} = \frac{\Delta_{LT}}{k_{LT}^2}  \left( i \frac{\partial}{\partial x} \pm  \frac{\partial}{\partial y} \right)^2
\end{equation}
with $\Delta_{LT}$ being half the TE-TM splitting at wavevector $k_{LT}$. The strength of the TE-TM splitting is defined by the ratio $\Delta_{LT}/k_{LT}^2$, while the in-plane wavevector of polaritons is given by the operator in the round brackets.
In all the theoretical calculations the following parameters were set to: $\alpha = 2.4$ $\mu$eV $\mu$m$^2$, $g_R = 1.5 \alpha$, $G = 4 \alpha$, $r_c = 0.01$ $\mu$m$^2$ ps$^{-1}$, $\Delta_{LT}/k_{LT}^2 = 11.9$ $\mu$eV $\mu$m$^2$, $\tau_p = 3.8$ ps, $\tau_x = 10$ ps.
%
%
%
%
\section{Discussion}
\label{sec:4}

\subsection{Experimental Results}
\label{subsec:a}

We investigate the formation mechanism of the spin textures under linearly polarized pulsed excitation. The nonresonant excitation creates electron-hole pairs in the QWs, which rapidly relax in energy toward the high $k$-states of the exciton dispersion, giving rise to the exciton reservoir \cite{Piermarocchi_exciton_1997}. The lower polariton dispersion is populated through exciton-phonon and exciton-exciton scattering \cite{kavokin_microcavities_2007}. 
\\
Here, we are interested in studying the pseudospin properties of polaritons which are directly related to the polarization of the emitted light by means of the Stokes vector \cite{kavokin_microcavities_2007}. By performing polarization resolved measurements and using the tomography technique described in Sec.~\ref{sec:2}, we time-resolve the polariton emission and observe their spin dynamics in real space. A summary of the experimental data taken for the specific excitation energy of 1.687 eV and wavevector $k$ $\leq\unit[2.9]{\mu m^{-1}}$ is shown in Figs.\ref{fig:1}(a-l). The linear [Figs.\ref{fig:1} (a-d)], diagonal [Figs.\ref{fig:1} (e-h)] and circular [Figs.\ref{fig:1} (i-l)] components of the Stokes vector, calculated by applying Eq.\ref{eq:Stokes_eq}, are shown at times $\unit[4]{p s}$, $\unit[14]{p s}$, $\unit[24]{p s}$ and $\unit[34]{ps}$ upon relaxation. The theoretical simulations realized with the model and the parameters described in Sec.\ref{sec:3} are shown in Figs.\ref{fig:1} (m-o).
\\
Under non-resonant excitation, the blueshift of polaritons is mainly determined by the interaction with the exciton reservoir \cite{Piermarocchi_exciton_1997, Pieczarka_Ghost_branch_2015}. In a recent work, we have shown how the exciton-exciton interactions in the proximity of the excitation spot directly affect the spin dynamics of polaritons, giving rise to a rotation of the circularly polarized spin textures, i.e., polariton spin whirls \cite{spin_whirls_2015}. The whirling of the spin textures is a consequence of a spin imbalanced exciton reservoir, which results in a splitting $g_R(\mathcal{N}_+ - \mathcal{N}_-)$ of polaritons acting as an effective magnetic field along the $z$-direction \cite{spin_whirls_2015}. 
In the current work, we use a lower excitation density compared to the spin whirls case \cite{spin_whirls_2015} and explore the regime where the exciton density and consequently the splitting of the exciton reservoir $g_R(\mathcal{N}_+ - \mathcal{N}_-)$ is not strong enough to cause any significant dynamic rotation in the polarization of the polariton condensate. 
\subsection{Skyrmion Number}
\label{subsec:b}

The spin texture of a skyrmion is characterized by a winding number, known as the skyrmion number $N_{sk}$, which is defined by the surface integral
\begin{equation}
N_{sk} = \frac{1}{4\pi}\int\textbf{S}\cdot\left(\frac{\partial \textbf{S}}{\partial x}\times\frac{\partial \textbf{S}}{\partial y}\right)dx dy.
\label{eq:Sk_number}
\end{equation}
Physically, it counts how many times $\textbf{S}$ wraps around the unit sphere when the integral covers the vortex core \cite{nagaosa_topological_skyrmion_2013}. This winding number is conserved and skyrmion is topologically non-trivial whenever the boundary condition of $\textbf{S}$ is fixed, e.g., by energetics~\cite{Lovegrove_energetic_Stability_skyrmions_2014}.
\\
In the case of polariton microcavities, this vector corresponds to the total pseudospin vector (see Eq.1 of the SI, for the full analytical expression of $\textbf{S}$ in polar coordinates $r$ and $\theta$), and its magnitude represents the total degree of polarization. Thus, by plotting the total pseudospin vector $\textbf{S}$ in the microcavity $x$-$y$ plane, the topological structure of the polariton half-skyrmion textures can be visualized. This is shown in Fig.~\ref{fig:1b}(a) for the first quadrant, color-coded with the three stokes components $S_x$,$S_y$,$S_z$. The regions where the total pseudospin vector $\textbf{S}$ is perpendicular to the $x$-$y$ plane correspond to the two circular polarization domains, with $\textbf{S}$ pointing up for $\sigma_+$ (red) and down for $\sigma_-$ (blue) domains. In these domains, indicated for reference by the dashed black lines in Fig.~\ref{fig:1b}(b), the skyrmion number $N_{sk}$ is equal to $-0.5$.
\\
Spin textures with a half-integer $|0.5|$ topological charge correspond to \textit{half-skyrmion} (also known as Merons\cite{Leslie_2009} or Mermin-Ho vortices \cite{Mermin_vortices_1976}) spin textures. In a skyrmion with a topological charge $N_{sk}=\pm1$, the $\textbf{S}$ vector performs a $\pi$ rotation with respect to the $x$-$y$ plane, over the integration domain, e.g., $S_z$ rotates continuously from $\pm1$ to $\mp1$. In a half-skyrmion, on the other hand, the half integer topological charge ($N_{sk}=\pm0.5$) means that the total pseudospin vector performs only a $\pi/2$ rotation over the integration domain, e.g., the $S_z$ component goes from $\pm 1$ to zero (or vice versa) from the half-skyrmion core to its domain boundary. Consequently, the $\textbf{S}$ vector, initially pointing toward the north (south), lies in the $x$-$y$ plane at the boundary of the integration domain. 
The sign of the topological number $N_{sk}$ is determined first, by the rotation of $S_z$ from the half-skyrmion core towards the integration boundary. Secondly, it is determined by the rotation of the in-plane spin ($S_x,S_y$) along a closed path containing the half-skyrmion core. In Fig.~\ref{fig:1b}(a), for example, $N_{sk}$ retains the value $-0.5$ from lobe to lobe in the same quadrant since $S_z$ and ($S_x,S_y$) both switch rotations between the half-skyrmion domains. Indeed, as shown in Fig.\ref{fig:total_spin} of the SI, $N_{sk}$ only changes sign between the quadrants of the system, a consequence of the OSHE. An analytical derivation showing the half-integer nature of the half-skyrmions is given in the SI.
\\
It is worth noting that differently from spinor atomic condensates, where the nonlinear interactions within the condensate are important for the stability of the spin textures \cite{Lovegrove_energetic_Stability_skyrmions_2014}, here the half-skyrmions appear in the expansion of the condensate where polariton-polariton interactions do not play a significant role. It is the spatial potential profile of the polariton condensate that determines the formation of half-skyrmions (see section \ref{subsec:d}) and fixes the asymptotic orientation of the spins outside their cores, thus ensuring their topological stability.

\subsection{Optical Spin Hall Effect}
\label{subsec:c}

The polariton pseudospin dynamics is mainly determined by the TE-TM splitting of the photonic modes (L-T splitting) \cite{maragkou_optical_2011,No_Solitons_2014}. At $k>0$, polaritons are split into two nondegenerate modes with polarization along (L) and orthogonal (T) to $\mathbf{k}$ and frequencies $\omega_L(k)$ and $\omega_T(k)$, respectively, with the LT splitting $\Delta_{LT}$. This splitting acts as a wavevector-dependent effective magnetic field ($\mathbf{H}_\text{LT}$) in the plane of the microcavity ($x$-$y$ plane) making the pseudospin of polaritons precess (see SI, section \ref{sec:8}), similar to the Rashba field in the case of electron spin in doped QWs \cite{Sinova_Intrinsic_SHE_2004}. The effective magnetic field $\mathbf{\Omega}_{\mathbf{k}}$ lies in the plane of the microcavity and its components are \cite{kavokin_optical_2005}:
\begin{equation}
\Omega_x = \frac{\Delta_{LT}}{\hbar k^2}(k_x^2 - k_y^2),\qquad \Omega_y = \frac{\Delta_{LT}}{\hbar k^2}\,2 k_x k_y,\qquad \Omega_z=0.
\label{eq:in_plane_components}
\end{equation}
Here, $\mathbf{k} = (k_x, k_y)$ is the in-plane polariton wave vector.
\begin{figure}[!hbtp]
  \centering
		\includegraphics[scale=0.5]{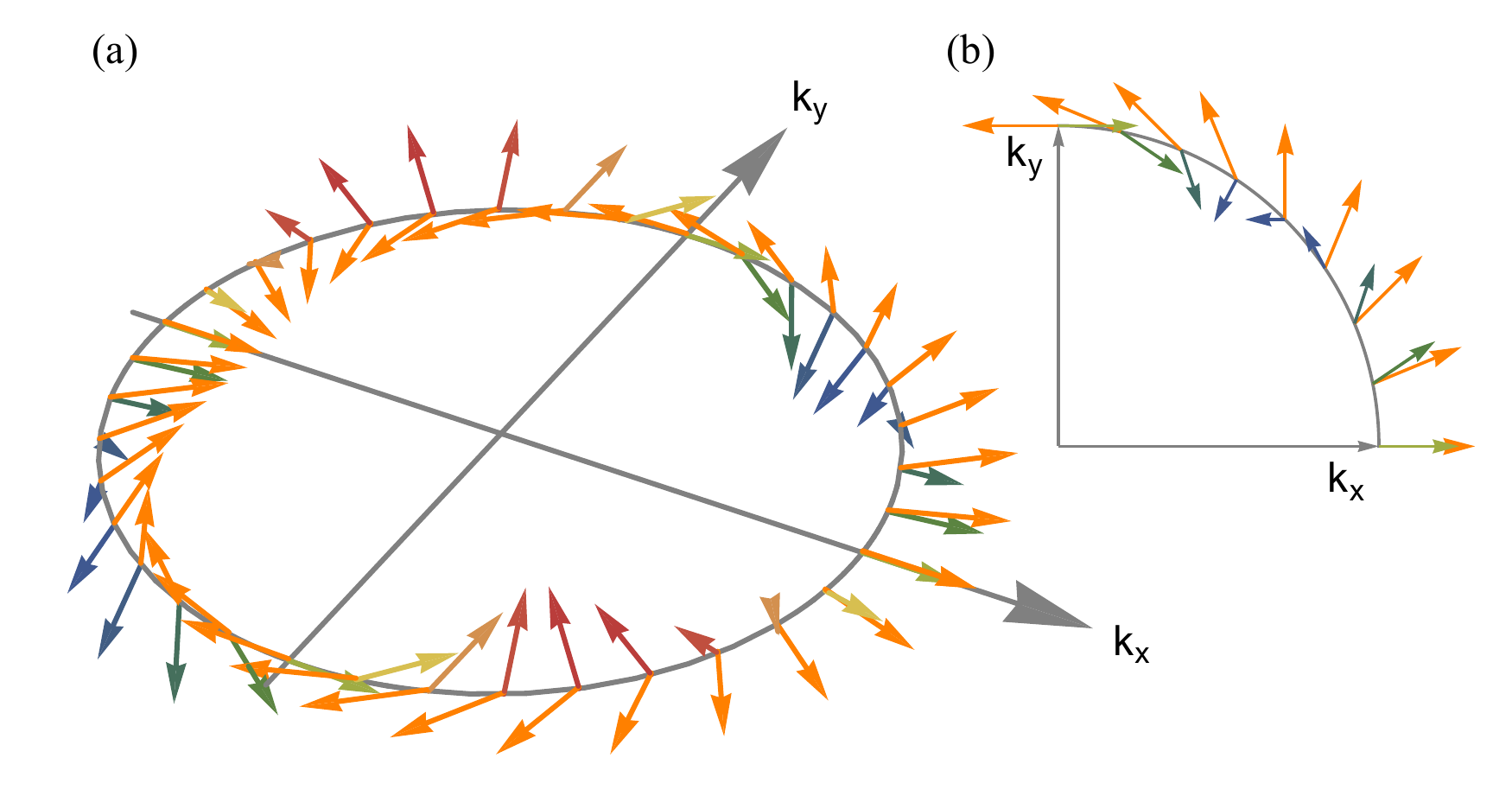}
	\caption{(Color online) (a) Sketch of the optical spin Hall effect in momentum space. The orange arrows show the effective magnetic field due to TE-TM splitting. The other arrows show the rotated polariton Stokes' vectors (starting from a linearly polarized state). The inset (b) shows the projections in the $x$-$y$ plane. Note that in any particular quadrant in reciprocal space, the sign of the $y$-component of the Stokes' vector reverses sign. This behavior is at the origin of the eight-lobe textures observed in the diagonal Stokes pattern [see Figs.~\ref{fig:1}(g,h) and Fig.~\ref{fig:1}(n)].}
\label{fig:2}
\end{figure}
\\
The magnitude of the effective magnetic field is proportional to $\Delta_{LT}$ and its direction in the plane of the microcavity depends on the direction of the polariton wave vector ($k_x$,$k_y$). As shown in Fig.~\ref{fig:2}, different values of the wavevectors correspond to a different orientation of the effective magnetic field in $k$-space (orange arrows), which in turn corresponds to a different precession of the pseudospin (colored arrows).
In the case of GaAs microcavities this effective magnetic field is at the origin of the OSHE \cite{kavokin_optical_2005}, which resembles the spin Hall effect in semiconductor thin layers \cite{dyakonov_spin_Hall_1971}. In the spin Hall effect, however, initially unpolarized electrons spontaneously separate in spin up and spin down fractions due to the electron spin-orbit interactions, while in the OSHE the initial polarized polaritons rotate their pseudospin due to an effective magnetic field, analog of the spin-orbit interaction.
\\
The OSHE, essentially consists in the angular polarized emission of the polaritons resulting in the appearance of alternating circularly or linearly polarized domains in the plane of the microcavity. The orientation of the spin polarized domains is determined by the polarization of the pump. For example, in a previous work we have shown that under circularly polarized pump, the highly imbalanced spin population driven by the pump \cite{Hamid_2012} results in concentric ring patterns of opposite circularly polarization \cite{kammann_nonlinear_2012}.
Here, differently from Ref.~[\onlinecite{kammann_nonlinear_2012}], we excite our sample with a linearly polarized beam and observe skyrmionic textures theoretically predicted for atomic \cite{al_khawaja_skyrmions_2001}, indirect excitons \cite{Skyrmion_Indirect_exciton_2013}, polariton condensates \cite{flayac_transmutation_2013} and the polarization beats experimentally observed in a microcavity under pulsed excitation \cite{langbein_polarization_2007}. In this case, the linearly polarized excitation results in a linearly polarized condensates, as shown in Figs.~\ref{fig:3}(a,c). A small imbalance between the two circular polarization components still persists [Fig.~\ref{fig:3}(c)], as a consequence of the small ellipticity introduced by the high-NA excitation objective used in the experiment \cite{spin_whirls_2015}. At high excitation densities this imbalance in the exciton reservoir, $g_R(\mathcal{N}_+ - \mathcal{N}_-)$, causes the rotation of the spin textures \cite{spin_whirls_2015}. Here, due to the low excitation density regime the spin textures in the microcavity plane remains fixed in time (i.e., they do not rotate) [Figs.~\ref{fig:1}(i-l)], making the observed effect essentially linear (i.e., with negligible polariton-polariton interactions).
%
\begin{figure}[!hbtp]
  \centering
		\includegraphics[scale=0.5]{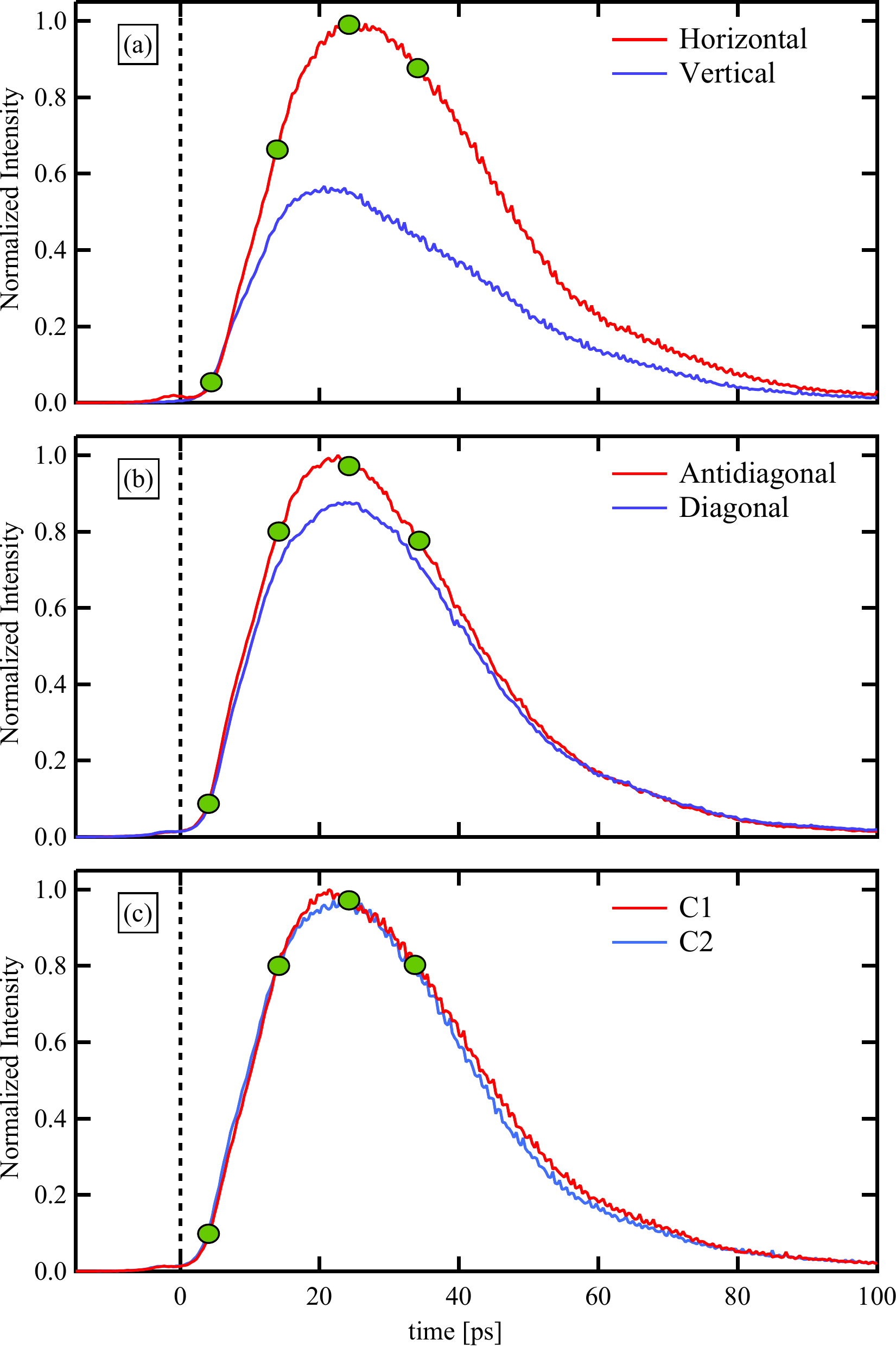}
		\caption{(Color online) Time-resolved, spatially integrated measurements of the (a) linear (b) diagonal and (c) circular polarization components photoluminescence intensity, normalized and integrated over over the area imaged in Figs.~\ref{fig:1}(a-l), i.e., $\unit[(400 \times 300)]{\mu m^2}$. The zero time is defined at the PL onset, as shown in the graphs. The green circles correspond to the time of the snapshots shown in Figs.\ref{fig:1}(a-l).}
\label{fig:3}
\end{figure}

\subsection{Formation dynamics of half-skyrmion spin textures}
\label{subsec:d}

Polaritons are generated nonresonantly by means of a tight focused spot of $\sim \unit[2]{~\mu m}$ FWHM. Due to the interaction with the exciton reservoir, polaritons are radially expelled out of the excitation spot. As they propagate outside of the excitation spot, the potential energy is converted to kinetic energy, with wavevector determined by the gradient of the potential induced by the blueshift of the condensate. Depending on the wavevector, polaritons propagating in different directions experience different polarization rotation due to $k$-dependent precession of the polariton pseudospin around the effective magnetic field $\mathbf{H}_{LT}$. Consequently, the formation of the half-skyrmion spin textures in the plane of the microcavity is angle dependent [Figs.~\ref{fig:1}]. In fact, for both linear [Figs.~\ref{fig:1}(a-d)] and circular [Figs.~\ref{fig:1}(i-l)] Stokes components, the polarization shows maxima in the diagonal direction while it is almost suppressed in the vertical and horizontal direction, reproducing the polarization quadrature typical of the OSHE\cite{kavokin_optical_2005}. Specifically, the horizontal and vertical directions correspond to the position where the effective magnetic field is parallel or antiparallel to the pseudospin, thus no precession occurs (see for example Fig.~\ref{fig:1}(d)). On the other hand, the diagonal directions (i.e., at $45^{\circ}$ respect to the $x$-$y$ axis) correspond to the directions where the pseudospin precesses around a perpendicular oriented effective magnetic field, giving rise to spin textures that appear as domains of opposite polarization as polaritons propagate radially out of the excitation spot.
Under CW excitation (see SI, Section \ref{sec:9}) the generation of polaritons, sustained by the continuous injection of electron-hole pairs due to the nonresonant CW laser, allows us to observe the polariton spin precess twice with respect to its initial orientation. This corresponds to the appearance of external polarized lobes in real space [Figs.S4(a) and (b) of the SI].
\\
In addition to the linear and circular spin textures, here we show the diagonal Stokes component of the condensate, as shown in Figs.~\ref{fig:1}(e-h). A characteristic eight-lobes textures centered around the excitation spot at $\unit[(0,0)]{\mu m}$ is observed [Fig.~\ref{fig:1}(g) and (h)]. The formation of this spin texture is due to the symmetry of the TE and TM states over the elastic circle in the ($k_x,k_y$) plane. In particular, the angle between the effective magnetic field ($\mathbf{H}_{LT}$) and the polariton wave vector corresponds to a double angle ($2\phi$) with respect to the $x$-axis in the Poincar\'e sphere. As sketched in Fig.~\ref{fig:2}, at any particular quadrant in reciprocal space, the sign of the $y$-component of the Stokes' vector reverses sign. Consequently, in each quadrant in $k$-space (inset in Fig.~\ref{fig:2}) there will be two opposite projections of the Stokes vector, which will correspond to the two opposite diagonal polarized lobes (for each quadrant) in real space [Figs.~\ref{fig:1}(g-h)].
\\
Theoretical simulations performed in the presence of disorder show that the half-skyrmion spin textures are stable against perturbations, such as structural defects naturally present in microcavities (see SI, section \ref{sec:10}).

\section{Conclusions}
\label{sec:5}

In conclusion, we have studied the formation and time evolution of two-dimensional half-skyrmion spin textures in planar semiconductor microcavities. We have demonstrated theoretically and experimentally that the appearance of these nontrivial spin textures is due to the optical spin Hall effect, which originates from the TE-TM splitting of the propagating modes. We note that the major contribution to the observations reported here is due to the splitting of the cavity optical modes which, consequently, makes the effect essentially linear (i.e., with negligible polariton-polariton interactions). The calculation of the characteristic skyrmion number, associated with the topology of half-skyrmion spin textures, supports and completes the observation.
\\
Vectorial textures with different topological charges have been studied in several physical systems, ranging from polarized optical beams \cite{dennis_polarization_2002}, semiconductor lasers \cite{Prati_VCSEL_1997} and atomic spinor condensates \cite{Hansen_singular_atom_optics_2016}. The study of these topological objects helps to better clarify the link between different branches of physics and to gain an in depth understanding of other physical systems \cite{analogies_power_2014}. 
Compared to conventional condensed matter systems, polariton condensates in semiconductor microcavities provide a unique opportunity to study and characterize spinor dynamics since the condensate order parameter can be directly accessed through optical measurements in both real and momentum space. Moreover, depending on the polarization of the excitation pump, several spin textures can be realized (see for example Ref.~[\onlinecite{kammann_nonlinear_2012}]), making polariton microcavities a suitable system to envisage a deterministic control of the skyrmionic spin textures by external optical beams.

\section*{ACKNOWLEDGMENTS}
\label{sec:7}
P.C., S.H. and P.L. acknowledge support by the Engineering and Physical Sciences Research Council of UK through the ``Hybrid Polaritonics'' Program Grant (Project EP/M025330/1). H.S. and I.A.S acknowledge the support from Rannis projects BOFEHYSS and Singaporean Ministry of Education under AcRF Tier 2 grant MOE2015-T2-1-055. I.A.S. thanks 5-100 program of Russian Federal Government. All data supporting this study are openly available from the University of Southampton repository at \url{http://dx.doi.org/10.5258/SOTON/386411}.
%
%

\newpage
\begin{widetext}

\vspace{2.0cm}

\begin{center}
{\large\textbf{Supplementary Information}}
\end{center}

\vspace{1.0cm}

\section{Experimental setup}
\label{sec:7a}

To conduct the experiments we use the experimental setup shown in Fig.~\ref{fig:S1}. We perform both time resolved and time integrated experiments. In the following we describe only the time integrated measurements reported in Fig.~\ref{fig:4}. The time resolved measurements [Figs.~1(a-l)] have been described in the main manuscript (see Section II, ``Sample and Experiment'' for details).  
\begin{figure}[!hbtp]
\includegraphics[scale=0.6]{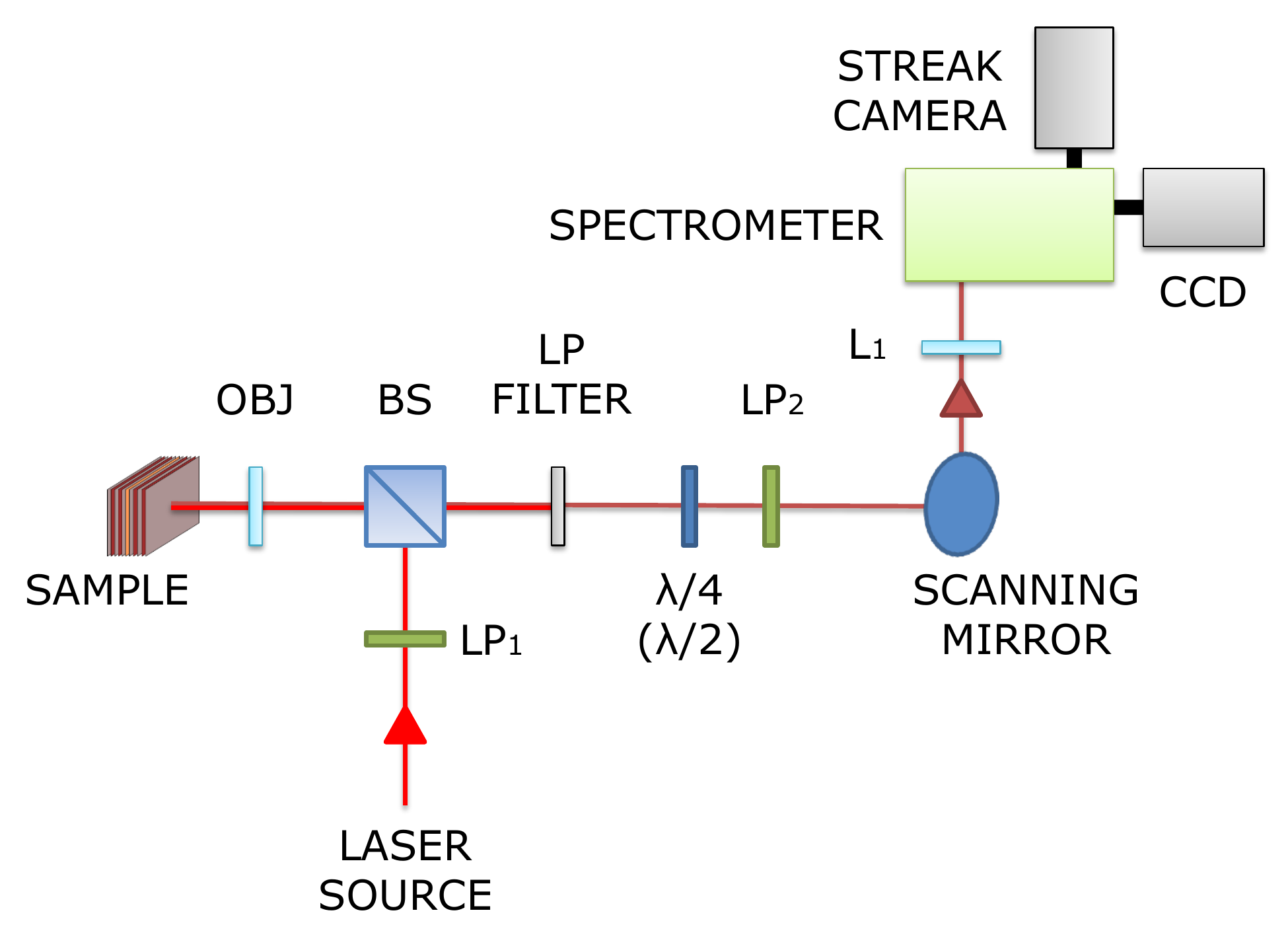}
\caption{Sketch of the setup used in the experiments. Lists of the optical components: \textbf{OBJ} is the 20x, 0.4 NA objective; \textbf{BS} is the non-polarizing beam splitter; \textbf{LP FILTER} is the long pass filter to filter the excitation laser out; \textbf{$\lambda/4~(\lambda/2)$} is the quarter-wave (half-wave) plate; \textbf{LP$_{1,2}$} are the linear polarizers and $\textbf{L1}$ is the $\unit[10]{cm}$ focal length lens.}
\label{fig:S1}
\end{figure}
%
%

In the time integrated experiments (Figs.~\ref{fig:4}), the excitation is provided by a single-mode narrow-linewidth CW laser, chopped with a duty cycle of 0.1 to reduce sample heating. The horizontal linearly polarized excitation, with polarization parallel to the $x$-axis (see Figs.~\ref{fig:4}), is tuned to the first reflectivity minimum above the stopband of the DBR, at $\unit[1.687]{eV}$ (i.e., non resonant excitation), and focused to a $\sim\unit[2]{~\mu m}$ FWHM spot by a 0.4 numerical aperture objective. The sample is held in a cold-finger cryostat at a temperature of $T \approx 6$ K. The polarized emission is collected in reflection geometry through the same objective and analyzed by a polarimeter composed of a $\lambda/2$ or $\lambda/4$ plate and a Wollaston prism, with $\sim$ 20$^{\circ}$ polarization splitting angle. The emission is then imaged in real space by a 10 cm focus lens directly on a CCD camera (time integrated measurements). In this way, both the polarization components are imaged simultaneously and the Stokes parameters calculated in real time by an appropriate software. 

\section{Total half-Skyrmion textures}
\label{sec:7b}

In Fig.~2 of the main manuscript we have shown the vector field of the total pseudospin vector $\textbf{S}$ calculated over the first quadrant. In Fig.~\ref{fig:total_spin}(a) we report the vector field of the total pseudospin vector $\textbf{S}$ in all the quadrants of the $x$-$y$ plane, together with the circular stokes component $S_z$ in Fig.~\ref{fig:total_spin}(b), showing the half-skyrmion topological indices calculated by means of the integral (Eq.5) of the article. The vector field of the total pseudospin vector reveals the presence of an ordered pattern composed of half-skyrmions, with topological index $N_{sk}=\pm0.5$. 
\begin{figure}[!htbp]
	\centering
		\includegraphics[scale=0.4]{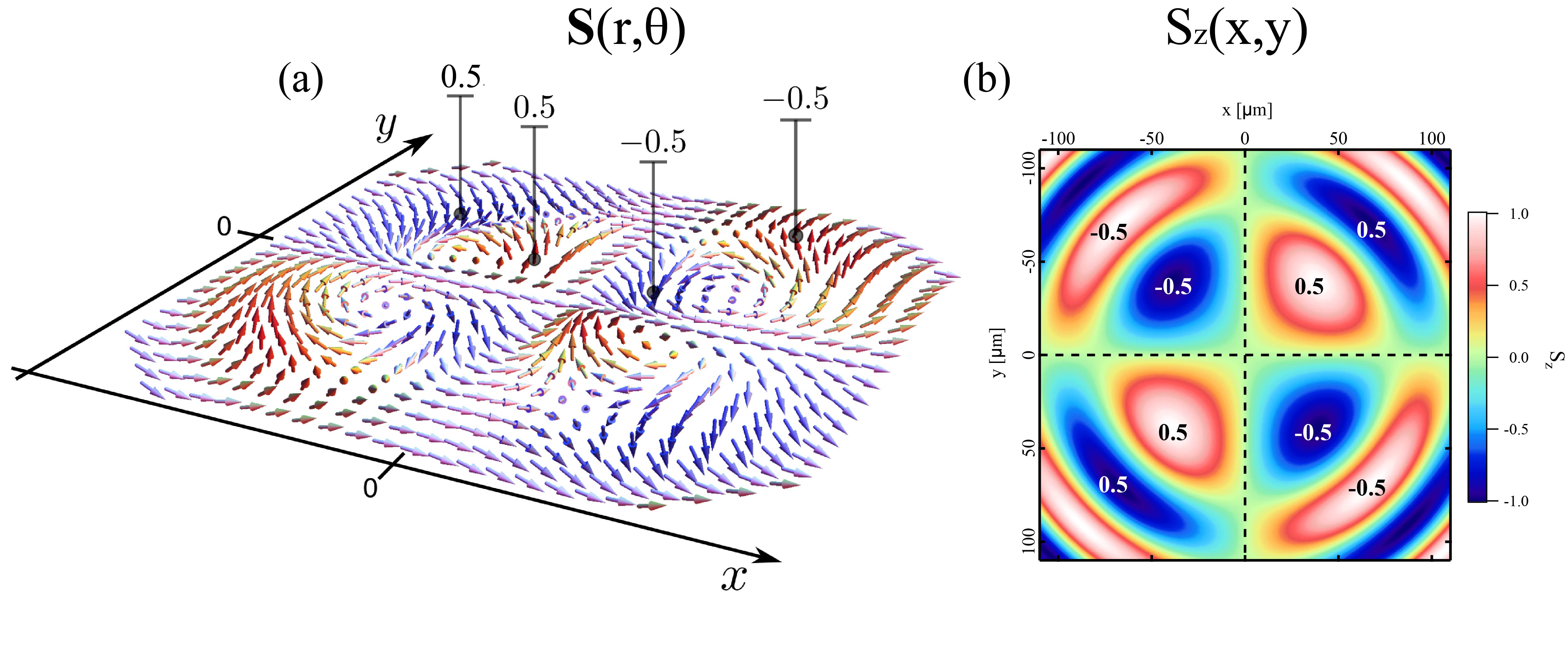}
\caption{(a) The vector field of the two-dimensional polariton half-skyrmions, showing the rotation of the total pseudospin vector $\textbf{S}(r,\theta)=(S_x,S_y,S_z)$ in the $x$-$y$ microcavity plane. The different colors of the vectors refer to the different polarization domains. In particular, red and blue refer to the circular polarizations $S_z$ (with opposite orientations along the z axis), while the other colors to the linear polarization components $S_x$ (lying in the $x$-$y$ plane), and the diagonal polarization $S_y$ (oriented at $\pm 45^{\circ}$ respect to the x-y plane). (b) Circular Stokes component $S_z$ showing the half-skyrmion topological indices calculated for each spin lobe by means of the integral (Eq.5) in the main manuscript.}
\label{fig:total_spin}
\end{figure}
\\
The sign of the topological number $N_{sk}$ is determined by the $S_z$ rotation from the half-skyrmion core towards the spin-lobe boundary, and the rotation of the in-plane pseudospin ($S_x,S_y$) along a closed path inside the lobe containing the half-skyrmion core. For example, the first lobe in the positive $x$-$y$ quadrant (colored blue) shows $S_z$ rotating from $-1$ to zero at the boundary of neighboring spin lobes. Around the half-skyrmion core, the in-plane pseudospin completes one counterclockwise rotation. Together, these give $N_{sk} = (-0.5) \times (1)$. In the next lobe (red), these have switched rotations and one gets $N_{sk} = (0.5) \times (-1)$.  

An approximate analytical expression of $\textbf{S}(r,\theta)$ for our non-resonant excitation can be associated with the solution of polaritons populating a single energy on a ring in $k$-space (see Ref.~[\onlinecite{flayac_transmutation_2013}]) where the polariton pseudospin field of the optical spin Hall effect can be described by:
\begin{equation}
\textbf{S}(r,\theta) = \left[\cos^2{(2 \theta)} + \sin^2{(2 \theta)} \cos{(\xi r)}\right]\hat{\textbf{x}}+\left[\sin{(4 \theta)}\sin^2{\left(\xi r/2\right)}\right]\hat{\textbf{y}}-\left[\sin{(2\theta)}\sin{( \xi r)}\right]\hat{\textbf{z}}
\label{eq:total_spin_vector}
\end{equation}
with $r$ and $\theta$ being the system polar coordinates and $\xi$ an arbitrary constant corresponding to the period of the $S_z$ spin rotating along the $x$-$y$ system diagonal. Plugging $\textbf{S}(r,\theta)$ into Eq.5 of the main manuscript results in:
\begin{align}	 \notag
N_{sk} = \frac{\xi}{4 \pi} \int \bigg[ & - 2 \cos^4{(2\theta)} \sin{(2\theta)} \sin^2{(\xi r)} \\ \notag
& + 4\cos^4{(2\theta)} \sin{(2\theta)} \cos{(\xi r)} \sin^2{ \left( \frac{\xi r}{2} \right)} \\ \notag
& - 4\cos^2{(2\theta)} \sin^3{(2\theta)}  \cos{(\xi r)} \sin^2{ \left( \frac{\xi r}{2} \right)} \\ \notag
& - 2\cos^2{(2\theta)} \sin^3{(2\theta)}  \cos{(\xi r)} \sin^2{ (\xi r)} \\ \notag
& + 4\cos^2{(2\theta)} \sin^3{(2\theta)} \cos^2{(\xi r)}  \sin^2{ \left( \frac{\xi r}{2} \right)} \\ \notag
&  - 4\sin^5{(2\theta)} \cos^2{(\xi r)}  \sin^2{ \left( \frac{\xi r}{2} \right)} \\ \notag
& - 4\cos^2{(2\theta)} \sin^3{(2\theta)}  \sin^2{(\xi r)} \sin^2{ \left( \frac{\xi r}{2} \right)} \\ \notag
& + 16 \cos^2{(2\theta)} \sin^3{(2\theta)} \cos{(\xi r)}  \sin^4{ \left( \frac{\xi r}{2} \right)} \\ 
& - 4 \sin^3{(2\theta)} \sin^2{(\xi r)} \sin^2{ \left( \frac{\xi r}{2} \right)} \bigg] dr d\theta 
\end{align}
Integration across individual spin lobes, $\theta \in [n  \to \  (n+1)] \pi/2$ and $r \in [m \to (m+1)]\pi/\xi$ with $n,m \in \mathbb{N}$, gives: 
\\
\begin{equation}
N_{sk} = \frac{(-1)^{n} }{60}  \bigg[ - 3  - 3 + 2 + 0 + 2  -8  -2 - 8 - 10 \bigg] = \frac{(-1)^{n+1} }{2},  
\end{equation}
\\
which shows the half-integer nature of the spin-lobes, confirming the presence of half-skyrmions. One can see that the sign of $N_{sk}$ only depends on which quadrant the spin lobe resides.
\\
\section{Optical Spin Hall Effect}
\label{sec:8}


One of the main effects affecting the spin dynamics of polaritons is the so called \textit{optical spin Hall effect} (OSHE). The OSHE, predicted by Kavokin and co-workers in 2005 \citep{kavokin_optical_2005} and experimentally observed in both polaritonic\,\cite{leyder_observation_2007} and photonic\,\citep{maragkou_optical_2011} microcavities, consists in the precession of the polariton pseudospin in the plane of the microcavity. The effect is enabled by the energy splitting between transverse-electric (TE) and transverse-magnetic (TM) polarized modes\,\cite{panzarini_exciton_1999} and the longitudinal-transverse splitting of the exciton states inside the microcavity \cite{maialle_exciton_1993}. The TE-TM splitting arises from the fact that different polarized optical modes will have different phase and penetrations into the Bragg mirrors. The splitting of the excitonic states, on the other hand, is mainly due to the long-range exciton exchange interaction and arises from the different alignment of the dipole moments (i.e., exciton states having dipole moments in different directions will have different energies \citep{kavokin_quantum_2004}). 
\newline
In the case of polariton microcavities, the TE-TM splitting ($\Delta_{LT}$) acts as a wave vector dependent effective magnetic field ($\mathbf{H_{eff}}$), making the pseudospin of polaritons precess if the latter is not parallel to it \citep{kavokin_optical_2005}. To take into account the precession of the pseudospin induced by the OSHE, polaritons propagation in microcavities is described by the following effective Hamiltonian \citep{kavokin_optical_2005}:
\begin{equation}
\widehat{H} = \frac{\hbar^2 k^2}{2 m^*} + \mu_B\,g (\bm{\sigma} \cdot \mathbf{H_{eff}}),
\label{eq:spin_Hamiltonian}
\end{equation}
where $m^*$ is the polariton effective mass, $\mu_B$ the Bohr magneton, $g$ the effective exciton Zeeman factor, $\bm{\sigma}$ the Pauli matrix vector and $\mathbf{H_{eff}}$ the effective magnetic field \citep{kavokin_optical_2005}: 
\begin{equation}
\mathbf{H_{eff}} = \frac{\hbar}{\mu_B\,g} \mathbf{\Omega_k}
\label{eq:effective_H}
\end{equation}
and $\mathbf{\Omega_k}$, which lies in the plane of the microcavity and has the following components \citep{kavokin_optical_2005}: 
\begin{equation}
\Omega_x = \frac{\Delta_{LT}}{\hbar k^2}(k_x^2 - k_y^2),\qquad \Omega_y = \frac{\Delta_{LT}}{\hbar k^2}\,2 k_x k_y,\qquad \Omega_z=0.
\label{eq:in_plane_components_SI}
\end{equation}
Here, $\stackrel{\rightarrow}{k}  = (k_x, k_y)$ is the in-plane polariton wave vector. As indicated by equation \ref{eq:in_plane_components_SI}, the orientation of the effective magnetic field in the plane of the microcavity, depends on the direction of the polariton wave vector, whose components are:
\begin{equation}
k_x = k\,\cos{\theta},\qquad k_y = k\,\sin{\theta},\qquad k_z=0,
\label{eq:wave_vectors}
\end{equation}
with $\theta$ being the angle between $\stackrel{\rightarrow}{k}$ (the direction of propagation) and the $k_x$-axis [see Figs.\ref{fig:arrows_k_space}]. Thus, by combining equation \ref{eq:in_plane_components_SI} and equation \ref{eq:wave_vectors}, the components of the effective magnetic field ($\mathbf{\Omega_k}$) can be determined\,\footnote{In the calculations the following trigonometric relations have been used: $cos^2(\theta) - sin^2(\theta) = cos(2\theta)$ and $2\,cos(\theta)\,sin(\theta) = sin(2\theta)$.}:
\begin{equation}
\Omega_x = \frac{\Delta_{LT}}{\hbar}\,\cos{(2\theta)},\qquad  \Omega_y = \frac{\Delta_{LT}}{\hbar} \sin{(2\theta)},\qquad \Omega_z=0
\label{eq:omega_components}
\end{equation}

Equation \ref{eq:omega_components} determines the orientation of the effective magnetic field in the plane of the microcavity, which depends on the direction of the polariton wave vector. The values of the effective magnetic field, calculated for different angles of propagation of polaritons, are reported in the table \ref{table:effective_m_field} (for the first and second quarter) and schematically represented in Fig. \ref{fig:arrows_k_space}(a). 
\begin{table}[ht]
\caption{Distribution of the effective magnetic field in the $k_x$-$k_y$ plane as function of the angle $\theta$, which defines the direction of propagation of polaritons [see Fig. \ref{fig:arrows_k_space}(a)].}
\centering 

\resizebox{\totalheight}{!}{\begin{tabular}{c c c} 	
	\hline\hline 
	
	Angle 									& $\Omega_x$ 										& $\Omega_y$ 										\\ [0.5ex] 
	
	\hline\hline  
	
$\theta=0$ 								& $\frac{\Delta_{LT}}{\hbar}$ 	& 0 														\\ [1ex]
	
$\theta=\frac{\pi}{4}$ 		& 0 														& $\frac{\Delta_{LT}}{\hbar}$ 	\\ [1ex]
	
$\theta=\frac{\pi}{2}$		& $-\frac{\Delta_{LT}}{\hbar}$	& 0 														\\ [1ex]
	
$\theta=\frac{3\pi}{4}$		& 0															& $-\frac{\Delta_{LT}}{\hbar}$	\\ [1ex]

$\theta=\pi$							& $\frac{\Delta_{LT}}{\hbar}$		& 0															\\ [1ex]

	\hline\hline  
	
\end{tabular}}
\label{table:effective_m_field} 
\end{table}
\begin{figure}[!htbp]
	\centering
		\includegraphics[scale=0.34]{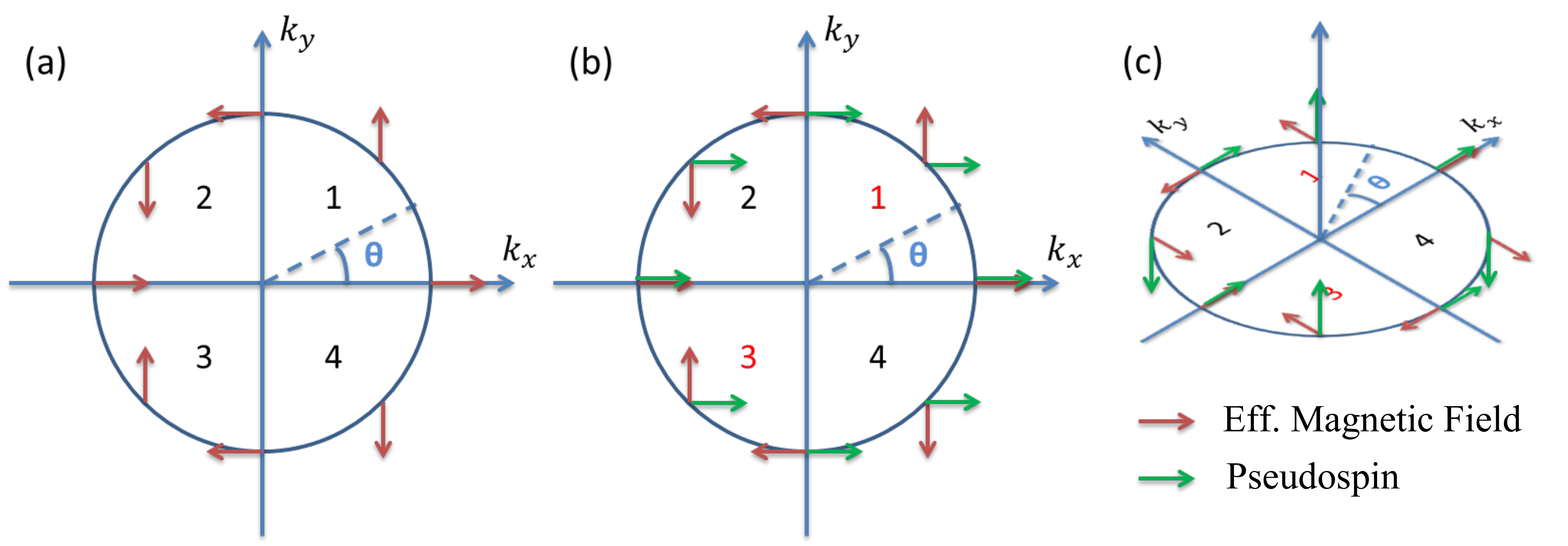}
\caption{(a) The red arrows show the distribution of the effective magnetic field in k-space induced by the TE-TM splitting (see equation \ref{eq:omega_components}). (b) The green arrows indicate the linearly polarized pseudospin (i.e., the pseudospin is parallel to the x-axis) while the red arrow correspond to the effective magnetic field. Note that at at $\theta = \pi/4,3\pi/4,5\pi/4,7\pi/4$ (i.e., along the diagonal directions with respect to the coordinates axes) the pseudospin is perpendicular to the effective magnetic field. (c) The initially linearly polarized pseudospin precess and, due to the orientation of the effective magnetic field, it becomes parallel to the z direction in the quarters 1 and 3 and antiparallel to the z axis in the quarters 2 and 4. Thus, the first quarters (1,3) correspond to  $\sigma_+$ while the other quarters (2,4) to $\sigma_-$ circularly polarized emission. Images adapted and redrawn from Ref.[\onlinecite{kavokin_optical_2005}].}
\label{fig:arrows_k_space}
\end{figure}
\newline
The TE-TM splitting of the polariton dispersion is zero at $\stackrel{\rightarrow}{k} =0$ and increases as a function of $k$, following a square root law at large $k$ \citep{kavokin_quantum_2004}. Since the magnitude of the effective magnetic field ($\mathbf{\Omega_k}$) is proportional to the TE-TM splitting ($\frac{\Delta_{LT}}{\hbar}$), also $\mathbf{\Omega_k}$ is zero at $k=0$.
By means of a tightly focused excitation spot, a radially expanding polariton condensate can be generated with a well defined wave vector. Since the orientation of the effective magnetic field depends on the polariton wave vectors (equation \ref{eq:in_plane_components_SI}), polaritons propagating in different directions experience different effective magnetic fields, which correspond to rotation of the pseudospin in different directions. In particular, polaritons propagating in opposite directions (i.e., at opposite angles $\theta$) experience precession in opposite directions [first and second quarter in Fig.\ref{fig:arrows_k_space}(b)]. This results to an angular dependent polarized emission of the polaritons and, consequently, to the appearance of alternating circularly or linearly polarized domains in the plane of the microcavity. Thus, different polarized spin domains develop in different quadrants of the $x$-$y$ plane [Fig.\ref{fig:arrows_k_space}(c)]. 
\\
\section{Spin Textures under CW excitation}
\label{sec:9}

We repeat the same experiment described in the main manuscript, but now exciting with a CW laser at power 5$\times P_{Thr}$ (with $P_{Thr}\approx15$ mW). 
\begin{figure}[!hbtp]
  \centering
		\includegraphics[scale=0.45]{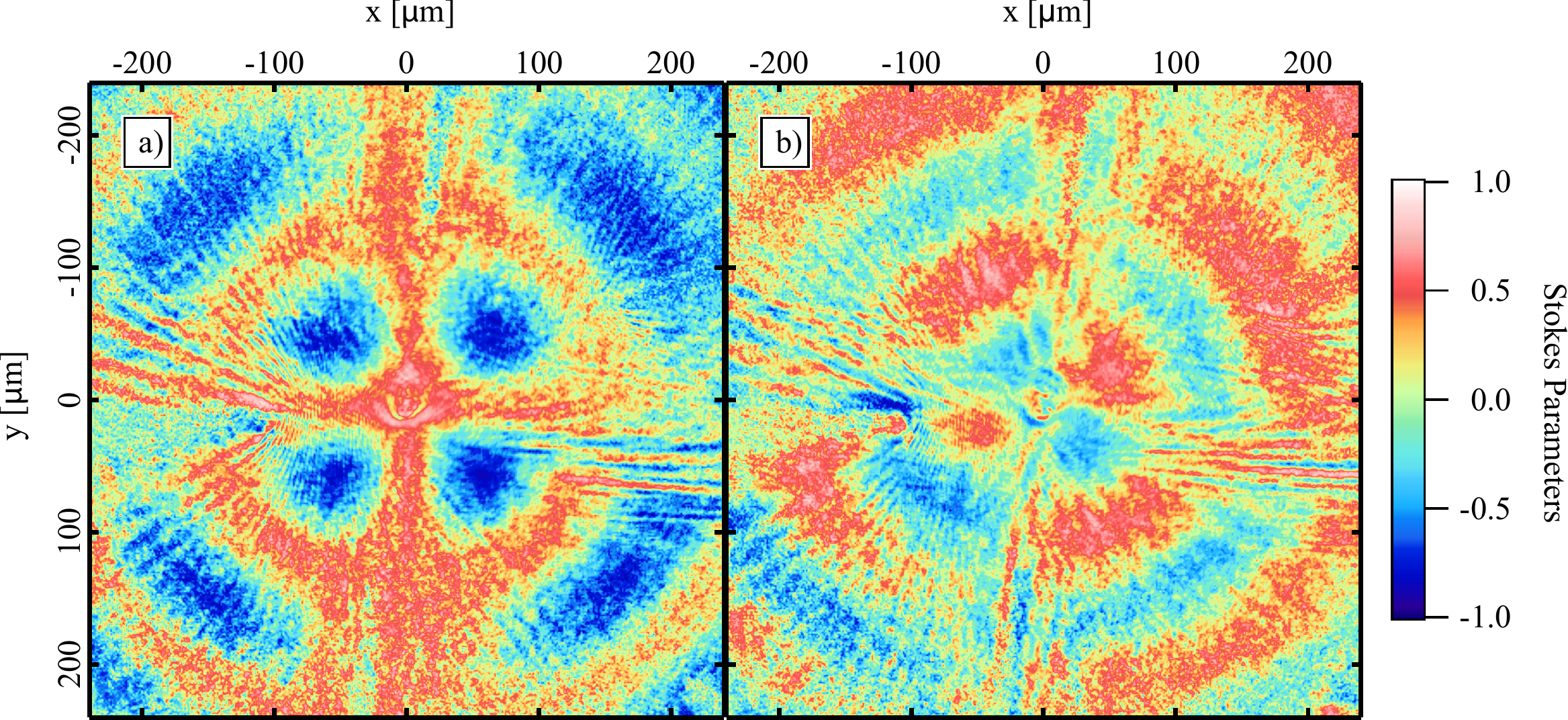}
	\caption{Experimental linear (a) and circular (b) Stokes parameters showing the formation of 2D pseudospin textures in real space. The excitation beam is linearly polarized and at $\unit[1.687]{eV}$.}
\label{fig:4}
\end{figure}

In CW experiments, polaritons decayed or emitted from the cavity are continuously replenished by the CW excitation so that, once a macroscopic ground state population is reached, i.e., the relaxation rate of polaritons into the ground state becomes greater than its radiative decay rate, a steady state polariton population is formed. In this case, the generation of polaritons is sustained by the continuously injection of electron-hole pairs due to the nonresonant CW laser, allowing us to observe the polariton spin to precess twice respect to its initial orientation [Figs.~\ref{fig:4}(a) and (b)]. This corresponds to the appearance of additional features (e.g., the external blue lobes in Figs.~\ref{fig:4}(a)) compared to the one observed in Figs.1(a-l) of the main manuscript.
\\
\\
\section{Spin Textures in presence of disorder}
\label{sec:10}

In Fig.~\ref{fig:S2}(a-c), the formation of the polariton spin textures shown in the main manuscript [Figs.1(m-o)] is calculated in the presence of disorder. The parameters used to perform the simulations are the same used for Figs.~1(m-o) of the main manuscript~\cite{parameters}. 
\begin{figure}[!hbtp]
\includegraphics[scale=0.6]{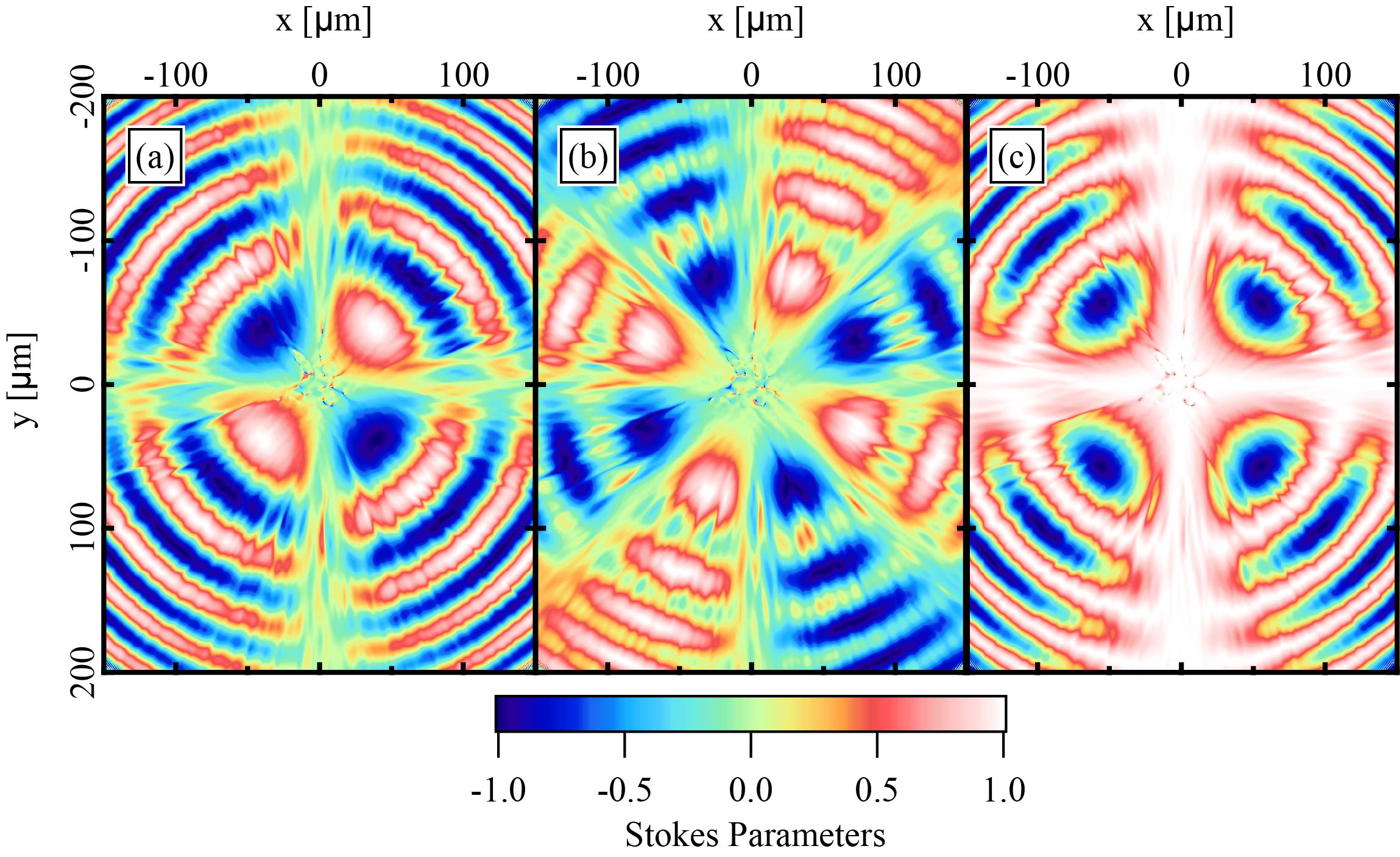}
\caption{Theoretical real space circular (a), diagonal (b) and linear (c) Stokes parameters calculated using Eqs.~2-4 from main manuscript in presence of disorder 34 ps after excitation with $P_+/P_- = 1$ and $P_+, \ P_- > 0$.}
\label{fig:S2}
\end{figure}

The disorder potential was generated with 0.05 meV root mean squared amplitude and 1.5 $\mu$m correlation length \cite{Savona_Langbein_2006}. The theoretical calculations show that disorder generates additional features in the spin textures, as in the case of the experimental data reported in Figs.\ref{fig:4}.
\end{widetext}
\bibliography{Bibliography}

\end{document}